\documentclass [runningheads]{svmult}
\usepackage{makeidx}
\usepackage{graphicx} 
\usepackage{subeqnar} 
\usepackage{multicol}
\usepackage{physprbb}
\makeindex

\begin{document}
\title*{Luttinger liquid behavior  in metallic carbon nanotubes}
\toctitle{Luttinger  liquid behavior in metallic carbon nanotubes}
\titlerunning{Luttinger liquid behavior in metallic carbon nanotubes}
\author{R.~Egger\inst{1}
\and A. Bachtold\inst{2}
\and M.S. Fuhrer\inst{2}
\and  M. Bockrath\inst{3}
\and  D.H. Cobden\inst{4} 
\and P.L. McEuen\inst{2}}
\authorrunning{R.~Egger et al.}

\institute{Fakult\"at f\"ur Physik, Universit\"at Freiburg, 
D-79104 Freiburg, Germany \and 
Department of Physics, 
University of California, Berkeley, CA 94720, USA
\and
Department of Physics, Harvard University, Cambridge, MA 02138, USA
\and
Department of Physics,
Warwick University, Coventry, CV4 7AL, UK}
\maketitle

\begin{abstract}
Coulomb interaction effects have pronounced consequences
in carbon nanotubes due to their 1D nature.
In particular, correlations imply the breakdown of Fermi liquid theory
and typically lead to Luttinger liquid behavior characterized by
pronounced power-law suppression of the transport current and the
density of states, and spin-charge separation.
This paper provides a review of the current
understanding of non-Fermi liquid effects in metallic single-wall
nanotubes (SWNTs).
We provide a self-contained theoretical discussion  of electron-electron
interaction effects and show that the tunneling
density of states exhibits power-law behavior. The power-law
exponent depends on the interaction strength parameter $g$ and on the
geometry of the setup.  
 We then show that these features are observed
experimentally by measuring the tunneling conductance of SWNTs as a
function of temperature and voltage. These tunneling experiments are
obtained by contacting metallic SWNTs to two nanofabricated gold
electrodes. Electrostatic force microscopy (EFM)
measurements show that the measured resistance is due to the contact
resistance from the transport barrier formed at the electrode/nanotube
junction. These EFM measurements show also the
ballistic nature of transport in these SWNTs.
While charge transport can be nicely attributed to
Luttinger liquid behavior, spin-charge separation has not been observed
so far.  We briefly describe a transport experiment that could provide
direct evidence for spin-charge separation.
\end{abstract}

\section{Introduction}

The electronic properties of one-dimensional (1D) metals have
attracted considerable attention for fifty years 
by now.  Starting with the work of Tomonaga 
in 1950 \cite{tomonaga} and later by Luttinger \cite{luttinger}, it has become
clear that the electron-electron interaction destroys 
the sharp Fermi surface and leads to a breakdown of the
ubiquituous Fermi liquid theory pioneered by Landau
\cite{landau}.   This breakdown is signalled by a vanishing
quasiparticle weight $Z_F$ in the presence of arbitrarily
weak interactions. The resulting non-Fermi liquid 
state is commonly called Luttinger liquid (LL), or sometimes
Tomonaga-Luttinger liquid.
The name ``Luttinger liquid'' was coined by Haldane \cite{haldane} to
describe the universal low-energy properties of one-dimensional
conductors.  Universality means that the 
physical properties do not depend on details of the model,
the interaction potential, etc.,  but instead are only characterized
by  a few parameters (critical exponents).  The range of
validity of the LL model is usually set 
by $E\ll D$, where $D$ is an electronic bandwidth parameter and
$E$ is the relevant energy scale, namely either the thermal scale $k_B T$
or the applied voltage $eV$. 
Quite remarkably, the  LL
concept is believed to hold for arbitrary statistical
properties of the particles, e.g.~both for fermions and
bosons.   It provides a paradigm for non-Fermi liquid physics
and may have some relevance also for higher-dimensional
systems, e.g.~in relation to high-temperature superconductivity.

In the model studied by Tomonaga and Luttinger, a special
dispersion relation for the noninteracting problem was
assumed, where one linearizes around the two Fermi vectors
$\pm k_F$ present in 1D.    At sufficiently low energy scales,
such a procedure should clearly be possible.  In fact, we will
see below that in a nanotube the dispersion relation is highly
linear anyways.   Assuming a linear dispersion relation composed
of left- and right-moving particles with Fermi velocity $v_F$,
one can equivalently express the noninteracting problem 
in terms of collective plasmon (density wave) excitations.
Technically, in the ``bosonization'' language \cite{bosonization},
for the simplest case of a spinless single-channel
system, these bosonic excitations   
can be expressed in terms of a displacement field $\theta(x)$
such that the density fluctuations are 
$\rho(x) = \pi^{-1/2} \partial_x \theta(x)$.
Electron-electron interactions then describe a bilinear coupling 
of these density fluctuations, and therefore
the full interacting problem can be written as a free theory 
in the displacement field:
\begin{equation} \label{h1}
H= \frac{\hbar v_F}{2} \int dx \left( \Pi^2(x) + \frac{1}{g^2} 
[\partial_x \theta(x)]^2 \right)
\;,
\end{equation}
where $\Pi(x)$ is the canonical momentum to the field $\theta(x)$.
In the long-wavelength limit, one can approximate the
 Fourier transform $\widetilde{V}(k)$
of the 1D interaction potential by a constant $V_0=\widetilde{V}(0)-
\widetilde{V}(2k_F)$, and the
dimensionless $g$ parameter in Eq.~(\ref{h1}) is given by 
\begin{equation}\label{g1}
g = [1+V_0/\pi \hbar v_F]^{-1/2} \;.
\end{equation}
Note that for repulsive interactions we always have $g<1$, with small $g$
meaning strong interactions.  
The limit $g=1$ describes the Fermi gas ({\sl not}\
a Fermi liquid),
and the limit $g\to 0$ leads to a classical Wigner crystal. 
The model (\ref{h1}) is equivalent to a set of harmonic
oscillators and can therefore be solved exactly.  The 
physical interpretation can be elucidated by the use of the
bosonization formula for the electron operator itself \cite{bosonization}. 
Thereby, the creation operator for a right- or left-moving electron
($r=R/L=\pm$) can be written in the form
\begin{equation}  \label{bos1}
\psi_r(x) \simeq \frac{1}{\sqrt{2\pi a}} \exp\Bigl(irk_F x  
+ir \sqrt{\pi} \theta(x) + i\sqrt{\pi} \int^x dx' \Pi(x') \Bigr) \;,
\end{equation}
where $a\approx 1/k_F$ is a lattice constant. 
Using this expression, it is a simple matter 
to show that the sharp $T=0$ Fermi surface
is smeared out for $g<1$, with interaction-dependent power laws
 close to $k_F$.
Physically, this is because the electron is an unstable particle and
 spontaneously
decays into collective plasmon modes.   
Including the spin-1/2 degree of
freedom, one finds that the spin and charge plasmons 
also decouple and moreover
propagate with different velocities $v_c\neq v_s$. 
 This phenomenon is called
{\sl spin-charge separation} and implies that the spin 
and charge degrees of freedom
of an electron brought into a LL will spatially separate. 
Note that in a Fermi liquid
$v_c=v_s$ and therefore this characteristic feature will not show up.
Spin-charge separation is intrinsically a dynamical phenomenon outside the
scope of thermodynamics.

An interesting and closely related issue 
concerns the fractionalized stable excitations 
of the LL. While it is easy to establish the spin-charge 
separation phenomenon
in the bosonic plasmon basis,  
the nature of the expected fundamental  ``quasiparticles'' with
fractional statistics, similar to the famous Laughlin quasiparticles 
in the fractional quantum Hall (FQH) effect, is less clear.
In a 1D Hubbard chain, which is known to be a realization of the
 LL at low temperatures,
well-defined spinon and holon excitations exist. 
For a spinless system, one can establish that quasiparticles scattered by a 
weak impurity potential have fractional charge $ge$ and a statistical angle 
$\pi g$ \cite{mpa}.
Remarkably, the fractional charge can have {\sl any} --
 even irrational -- value.
Furthermore, for the topology of a LL on a ring, 
a complete characterization
of the universal LL theory in terms of 
fractional-statistics quasiparticles 
has been provided recently \cite{pham}.
 
In view of this discussion, it is understandable that, for many decades,
experimentalists have attempted to find LL behavior. 
In the 1970s, the key interest was focused on quasi-1D organic chain
compounds \cite{voit}, where LL behavior is hard to establish because
of complicated 1D-3D crossover phenomena and additional phase transitions
into other states.  The interest was revived a few years ago, when
experimental observations of LL behavior for transport 
in semiconductor quantum wires \cite{tarucha,auslaender}
and for edge states in FQH bars \cite{milliken,chang} were reported.
Shortly after the theoretical prediction of LL
 behavior in metallic carbon nanotubes 
\cite{egger,egger2},
the to-date perhaps cleanest experimental 
observations of LL behavior were established
in transport experiments for 
single-wall nanotubes (SWNTs) \cite{bockrath,yao}.
The theory along with the experiments of Ref.~\cite{bockrath} 
will be presented below. 
By now, there are also several other theoretical proposals 
for probing the LL state
in {\sl bulk}\ systems, e.g.~by 
investigating the tunneling density of states (TDOS)
of a 3D metal in an ultra-strong magnetic field \cite{glaz}, or by studying
2D arrays of regularly stacked nanotubes \cite{kane}.

Carbon nanotubes were discovered in 1991 by Iijima \cite{tube91} and
have enjoyed exponentially increasing interest since then. 
The current status of the field has been summarized  
in a recent Physics World issue \cite{pworld}, see also Ref.~\cite{dekker}. 
Ignoring the end structure, one may think of a SWNT as a graphene sheet,
i.e.~a 2D honeycomb lattice made up of C atoms, that is
wrapped onto a cylinder, with typical radius of order 1-2 nm and length of
several microns.  Depending on the helicity of the wrapping, the resulting
SWNT is either semiconducting or metallic.  In our experimental setup
discussed in Sec.~\ref{sec3},  these two behaviors can be distinguished
as follows.  When the conductance $G$ of the 
tube is measured as a function of a gate voltage $V_g$, 
$G$ is virtually independent of $V_g$ for metal
tubes, while $G$ varies exponentially with $V_g$ for 
semiconducting tubes. The discussion in this paper is limited to
transport through {\sl metallic} SWNTs, where LL behavior can be expected.

\begin{figure}
\includegraphics[width=\textwidth]{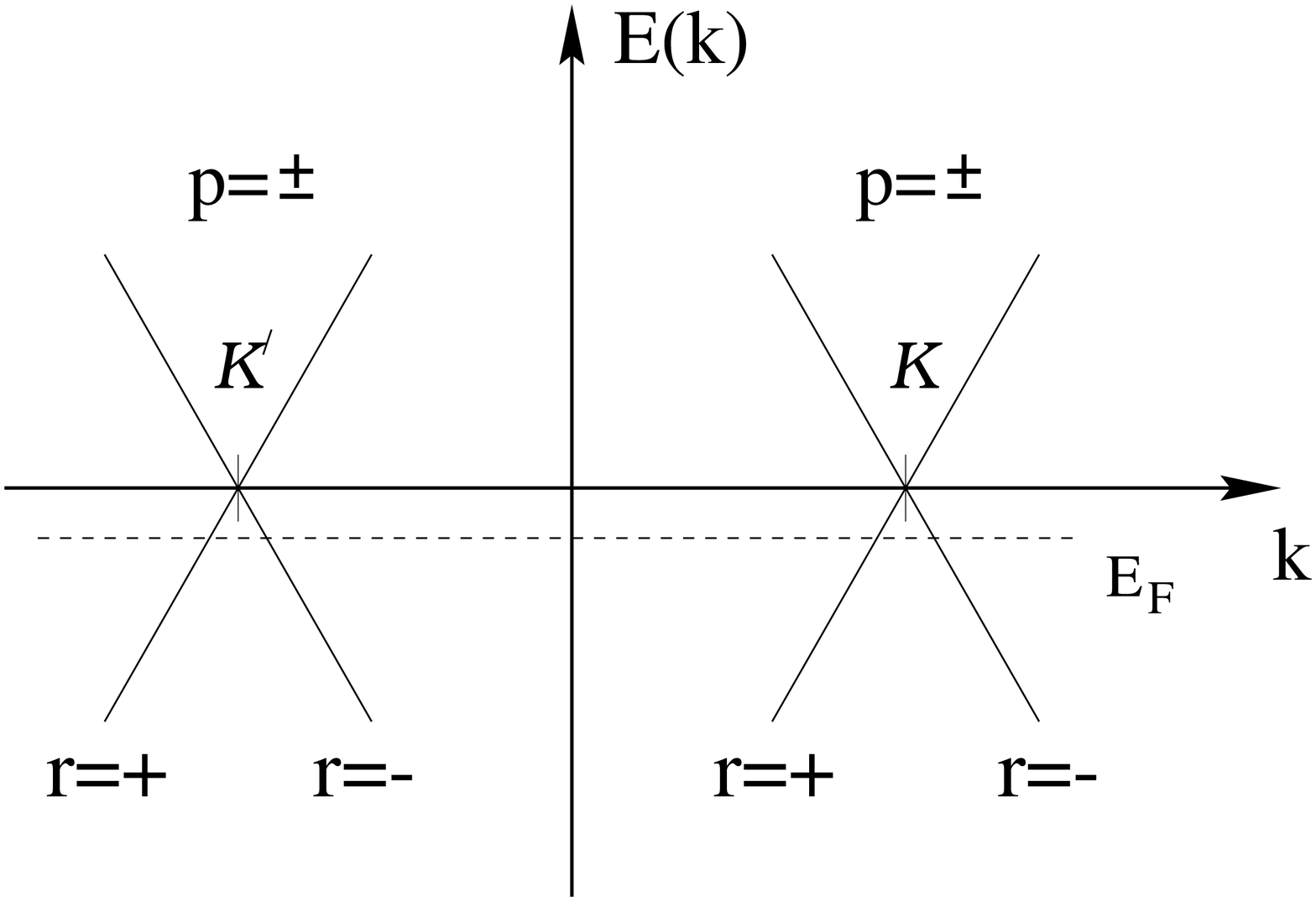}
\caption[]{\label{fig1} Schematic bandstructure of a metallic SWNT.
A right- and left-moving branch ($r=\pm$) is found near each of
 the two Fermi points $k=\alpha k_F$ with $\alpha=\pm$, corresponding to
$K$ and $K'$, respectively. Right- and left-movers arise as
linear combinations of the sublattices $p=\pm$. The Fermi energy
(dashed line) is shifted away from 
neutrality by doping and/or external gates.}
\end{figure}

$ $From the special band structure of a graphene sheet \cite{pworld}, 
one arrives at the characteristic dispersion relation of a metallic SWNT
shown in Figure \ref{fig1}.  This band structure exhibits two Fermi points
$\alpha=\pm$ with a right- and a left-moving ($r=R/L=\pm$) branch around
each Fermi point.  These branches are highly linear with  Fermi velocity
$v_F\approx 8\times 10^5$ m/s. The R- and L-movers arise as linear
combinations of the $p=\pm$ sublattice states reflecting the two C atoms
in the basis of the honeycomb lattice.  
The dispersion relation depicted 
in Fig.~\ref{fig1} holds for energy scales $E < D$, with
the bandwidth cutoff scale $D\approx \hbar v_F/R$ for tube radius $R$.  
For typical SWNTs,  $D$ will be of the order 1~eV.  The large overall energy 
scale together with the structural stability of SWNTs explain their
unique potential for revealing LL physics.  In contrast to conventional
systems, e.g.~semiconductor quantum wires, LL effects in SWNTs
are not restricted to the meV range but may even be seen at
room temperature.  An additional advantage is that
the approximation introduced by linearizing the dispersion relation 
in conventional 1D systems is here provided 
by nature in an essentially exact way.  A basic prerequisite of the
theory \cite{egger} is the {\sl ballistic} nature of transport in SWNTs.
Ballistic transport in SWNTs can be
unambiguously established by various experiments,
 see below and Ref.~\cite{tans,bockrath2,bachtold}.
Theoretical analysis \cite{white} has also suggested the absence of a 
diffusive phase in SWNTs, with the possibility of ballistic transport over
distances of several $\mu$m.
  
Besides SWNTs, LL effects have also been observed in the
TDOS of multi-wall nanotubes (MWNTs) \cite{schoenen,kim}.  MWNTs are composed
of several concentrically arranged graphene shells, and 
under the assumption of ballistic transport, the only incomplete screening
does not spoil the LL behavior \cite{mwnteg}.
On the other hand, transport in MWNTs has typical signatures
of {\sl diffusive} transport \cite{bachtold,schoenen}, and the theoretical
situation must be regarded as unsettled at the moment. 
We shall therefore only discuss (metallic)  SWNTs in this review.

The structure of the article is as follows.
In Sec.~\ref{sec2}, the theoretical description of a metallic SWNT in the
ballistic limit is reviewed, where we focus 
on the low-energy regime $E <  D$.
We shall derive the scaling forms of the nonlinear $dI/dV$ characteristics
for bulk or end tunneling into a nanotube, and point to various experimental
setups that can detect correlation effects in the transport.
We shall also briefly outline a recent suggestion for a spin-transport
 experiment that
could allow for the experimental verification of spin-charge separation.
In Sec.~\ref{sec3}, the experimental
evidence for LL behavior found so far is reviewed. 
Finally, in Sec.~\ref{sec4}, we summarize and discuss some of the
open problems that we are aware of.

\section{Luttinger-liquid theory for nanotubes}
\label{sec2}

\subsection{Low-energy theory: General approach}

The remarkable electronic properties of carbon nanotubes are
due to the special bandstructure of the $\pi$ electrons in
graphene. There are only two linearly independent Fermi points
$\alpha \vec{K}$ with $\alpha=\pm$ 
instead of a continuous Fermi surface. 
Up to energy scales $E < D\approx 1$ eV, 
the dispersion relation around the Fermi points 
is, to  a very good approximation, linear.
Since the basis of the honeycomb lattice contains two atoms,
there are two sublattices $p=\pm$, and hence  
two degenerate Bloch states 
\begin{equation}\label{bloch}
\varphi_{p\alpha}(\vec{r}) = 
(2\pi R)^{-1/2} \exp( -i\alpha \vec{K}\vec{r} ) 
\end{equation}
at each Fermi point $\alpha=\pm$.
Here  $\vec{r}=(x,y)$  lives
on the  sublattice $p$ under consideration, and
we have already anticipated
the correct normalization for nanotubes.
The Bloch functions are defined separately on each
sublattice such that they vanish on the other.
One can then expand the electron operator
in terms of these Bloch functions.
The resulting effective low-energy theory of
graphene is the 2D massless Dirac hamiltonian.
This result can also be derived 
in terms of $\vec{k}\cdot \vec{p}$ theory.

Wrapping the graphene sheet onto a cylinder then
leads to the generic bandstructure of a metallic SWNT
shown in Fig.~\ref{fig1}. 
Writing the Fermi vector as $\vec{K}= (k_F,p_F)$, where
the $x$-axis is taken along the tube direction and 
the circumferential variable is $0< y <2\pi R$,
quantization of transverse motion now allows for a
contribution $\propto \exp(i m y/R)$ to the wavefunction. 
However,  excitation of angular momentum states
other than $m=0$ costs a huge energy of order $D \approx 1$~eV.
In an effective low-energy theory, we may thus omit all transport
bands except $m=0$ (assuming that the SWNT is not
excessively doped).
Evidently, the nanotube forms a 1D quantum wire 
with only two transport bands 
intersecting the Fermi energy. This strict
one-dimensionality is fulfilled up to remarkably high
energy scales (eV) here, in contrast to conventional
1D conductors.
The electron operator for spin $\sigma=\pm$
is then written as
\begin{equation}\label{expa}
\Psi_\sigma(x,y) = \sum_{p\alpha} \varphi_{p\alpha}(x,y) 
\,\psi_{p\alpha\sigma} (x) \;,
\end{equation} 
which introduces slowly varying
 1D fermion operators $\psi_{p\alpha\sigma}(x)$ that depend 
only on the $x$ coordinate.  
Neglecting Coulomb interactions for the moment, 
the hamiltonian is: 
\begin{equation} \label{h0}
H_0= - \hbar v_F \sum_{p\alpha\sigma} p \int dx \;\psi_{p\alpha\sigma}^\dagger
\partial_x \psi^{}_{-p\alpha\sigma} \;.
\end{equation}
Switching from the sublattice
($p=\pm$) description to the right- and left-movers ($r=\pm$) 
indicated in Fig.~\ref{fig1} implies two copies ($\alpha=\pm$)
of  massless 1D Dirac hamiltonians for each spin direction.
Therefore a perfectly contacted and clean SWNT is expected to
have the quantized conductance $G_0=4e^2/h$.  Due to 
the difficulty of fabricating sufficiently good contacts,
however, this value has not been experimentally observed so far.
(We note that the conductance quantum $2e^2/h$ seen in recent MWNT
experiments by Frank {\sl et al.}~\cite{frank} is anomalous and 
does not correspond to the expected value of $G_0$.)
Remarkably, other spatial oscillation
periods than the standard wavelength $\lambda=\pi/k_F$ are possible. 
$ $From  Fig.~\ref{fig1} we observe  that
the wavelengths 
\begin{equation}\label{wavelength}
\lambda = \pi/k_F ,\quad \pi/|q_F|, \quad \pi/(k_F \pm q_F ) 
\end{equation}
could occur, where the doping determines the wavevector
 $q_F\equiv E_F/\hbar v_F$. Which of the wavelengths 
(\ref{wavelength}) is ultimately realized  sensitively depends
on the interaction strength \cite{egger2}. 

\subsection{Electron-electron interactions}

Let us now examine Coulomb interactions mediated by an arbitrary
potential $U(\vec{r}-\vec{r}')$. 
The detailed form of this potential will depend on properties of the
substrate, nearby metallic gates, and the geometry of the setup.
In the simplest case, bound electrons 
and the effects of an insulating substrate
are described by a dielectric constant 
$\kappa$,  and for an externally unscreened Coulomb interaction, 
\begin{equation}\label{unsc}
U(\vec{r}-\vec{r}') = \frac{e^2/ \kappa} 
{\sqrt{ (x-x')^2 + 4R^2 \sin^2[(y-y')/2R] + a_z^2 }}\;,
\end{equation}
where $a_z\approx a$ denotes the average distance between a
$2p_z$ electron and the nucleus, 
i.e.~the ``thickness'' of the graphene sheet.
We neglect relativistic effects like retardation or spin-orbit coupling
in the following.
Electron-electron interactions
are then described by the second-quantized hamiltonian
\begin{equation}\label{int0}
H_I = \frac12 \sum_{\sigma\sigma'}\int d\vec{r} \int
d\vec{r}' \, \Psi^\dagger_\sigma(\vec{r}) \Psi^\dagger_{\sigma'}
(\vec{r}') 
 U(\vec{r}-\vec{r}') \Psi^{}_{\sigma'}(\vec{r}')
\Psi^{}_\sigma(\vec{r}) \;.
\end{equation}
The interaction (\ref{int0}) can be reduced to a 
1D form by inserting the expansion (\ref{expa}) 
for the electron field operator.
The reason to do so is the large arsenal of theoretical
methods readily available for 1D models.
The result looks quite complicated at first sight:
\begin{equation}\label{general}
H_I= \frac12 \sum_{pp'\sigma\sigma'} 
\sum_{\{\alpha_i\}}
 \int dx dx'\; V^{pp'}_{\{\alpha_i\}}(x-x') 
\psi^\dagger_{p\alpha_1\sigma}(x) \psi^\dagger_{p'\alpha_2\sigma'}
(x') \psi^{}_{p'\alpha_3\sigma'}(x') \psi^{}_{p\alpha_4\sigma}(x) 
\;,
\end{equation} 
with the 1D interaction potentials
\begin{equation}\label{intpot}
V^{pp'}_{\{\alpha_i\}}(x-x') = \int dy dy'  
\varphi^{\ast}_{p\alpha_1}(\vec{r})
\varphi^{\ast}_{p'\alpha_2}(\vec{r}') 
U(\vec{r}-\vec{r}' + p \vec{d} \delta_{p,-p'} ) 
\varphi^{}_{p'\alpha_3}(\vec{r}') \varphi^{}_{p\alpha_4}(\vec{r}) \;.
\end{equation} 
These potentials only depend on $x-x'$ and on the 1D
fermion quantum numbers.
For interactions involving different sublattices $p\neq p'$ for $\vec{r}$
and $\vec{r}'$ in Eq.~(\ref{int0}), 
one needs to take into account the shift vector $\vec{d}$ between sublattices.

To simplify the resulting 1D interaction (\ref{general}), 
we now exploit momentum conservation, assuming $E_F\neq 0$
so that Umklapp electron-electron scattering can be ignored.
We then have  ``forward scattering'' processes,
where $\alpha_1=\alpha_4$ and $\alpha_2=\alpha_3$.
In addition, ``backscattering'' processes may
 be important, where $\alpha_1=-\alpha_2=
\alpha_3=-\alpha_4$.
We first define the potential
\begin{equation}\label{v0}
V_0(x-x')= \int_0^{2\pi R} \frac{dy}{2\pi R}
\int_0^{2\pi R} \frac{dy'}{2\pi R}\;  U(\vec{r}-\vec{r}')\; .
\end{equation}
For the unscreened Coulomb interaction  (\ref{unsc}), this
can be explicitly evaluated \cite{egger2}. 
$ $From Eqs.~(\ref{intpot}) and (\ref{bloch}),
the forward scattering  interaction potential reads
$V_0(x)+\delta_{p,-p'} \delta V_p(x)$,
with 
\begin{equation}\label{deltav}
 \delta V_p(x)  =  \int_0^{2\pi R} \frac{dy dy'}{(2\pi R)^2}
 [U(x+pd_x,y-y'+p d_y)- U(x,y-y') ] \;,
\end{equation}
which is only present if $\vec{r}$ and $\vec{r}'$ are
located on different sublattices.  Thereby important
information about the discrete nature of the graphite 
network is retained despite the low-energy continuum approximation.
Since  $V_0(x)$ treats both sublattices on equal footing,
the resulting  part of the forward scattering interactions couples only the
total 1D electron densities, 
\begin{equation}\label{fs0}
H_I^{(0)} = \frac12 \int dx dx' \, \rho(x) V_0(x-x') \rho(x') \;,
\end{equation}
where the 1D density is 
$\rho = \sum_{p\alpha\sigma} \psi^\dagger_{p\alpha\sigma}
\psi^{}_{p\alpha\sigma}$.  This part of the electron-electron
interaction is the most important one and will be seen to imply
LL behavior.  Note that it is entirely due to the
{\sl long-ranged}\, tail of the Coulomb interaction. 
All the remaining residual interactions come from short-ranged 
interaction processes, and since these are intrinsically averaged
over the circumference of the tube, their amplitude is quite
small and will (at worst) only cause exponentially small gaps.
A related general discussion can be found in Ref.~\cite{sawatzky}.

For $|x|\gg a$,  detailed analysis shows that $\delta V_p(x)=0$. However, 
for $|x|\leq a$, an additional term beyond Eq.~(\ref{fs0}) arises
due to the hard core of the Coulomb interaction. 
At such small length scales, the difference between
inter- and intra-sublattice interactions matters.
 To study this term, one should evaluate $\delta V_p(0)$  from
microscopic considerations.
One then finds the  additional forward scattering contribution \cite{egger2}
\begin{equation}\label{fs1}
H^{(1)}_I= 
- f \int dx\sum_{p\alpha\alpha'\sigma\sigma'}
\psi^\dagger_{p\alpha\sigma}\psi^\dagger_{-p\alpha'\sigma'}
\psi^{}_{-p\alpha'\sigma'} \psi^{}_{p\alpha\sigma} \;,
\end{equation}
where $f/a = \gamma_f e^2/R$.  An estimate  for armchair SWNTs 
yields $\gamma_f\approx 0.05$.
Since these short-ranged interaction processes are averaged
over the circumference of the tube, $f\propto 1/R$, and
hence $f$ is very small.
A similar reasoning applies to the backscattering contributions 
 $\alpha_1=-\alpha_2=\alpha_3=-\alpha_4$
in Eq.~(\ref{general}). 
Because of a rapidly oscillating phase factor, 
the only non-vanishing  contribution
comes again from $|x-x'|\leq a$, and we can effectively take a local 
interaction. Furthermore, only the
part of the interaction which does not distinguish among
the sublattices is relevant and leads to 
\begin{equation} \label{bs}
H_I^{(2)} = b \int dx\sum_{pp'\alpha\sigma\sigma'}
\psi^\dagger_{p\alpha\sigma}\psi^\dagger_{p'-\alpha\sigma'}
\psi^{}_{p'\alpha\sigma'} \psi^{}_{p-\alpha\sigma} \;.
\end{equation}
For the unscreened interaction (\ref{unsc}), 
$b/a = \gamma_b e^2/R$ with $\gamma_b\approx \gamma_f$.
For externally screened Coulomb interaction,  one may have
$b\gg f$.

Progress can then be made by employing the {\sl bosonization} 
approach \cite{bosonization}.
For that purpose, one first needs to bring the non-interacting
hamiltonian (\ref{h0}) into the standard form
of the 1D Dirac model.  This is accomplished by
switching to right- and left-movers
($r=\pm$) which are linear combinations of the sublattice
states $p=\pm$.
In this representation, a bosonization formula 
generalizing Eq.~(\ref{bos1}) applies, now 
with four bosonic phase fields $\theta_a(x)$ and their canonical momenta $\Pi_a(x)$.
The four channels are obtained from
combining charge and spin degrees of freedom as well as symmetric and antisymmetric linear
combinations of the two Fermi points, $a=c+,c-,s+,s-$.
The bosonized expressions for $H_0$ and $H_I^{(0)}$  read 
\begin{eqnarray} \label{bh0}
 H_0 &=& \sum_{a} \frac{\hbar v_F}{2} 
\int dx \left[  \Pi_a^2
+ g^{-2}_a (\partial_x \theta_a)^2 \right]\\
\label{bfs0}
H_I^{(0)} &=& \frac{2}{\pi} \int dx dx' \;
\partial_x\theta_{c+}(x) V_0(x-x') \partial_{x'} \theta_{c+}(x') \;.
\end{eqnarray}
The bosonized form of $H_I^{(1,2)}$ \cite{egger} leads to nonlinearities
in the $\theta_a$ fields for $a\neq c+$. 
Although bosonization of Eq.~(\ref{h0}) gives $g_a=1$ in
Eq.~(\ref{bh0}) [see also Eq.~(\ref{h1})],
interactions will renormalize these parameters.
In particular, in the long-wavelength limit,
 $H_I^{(0)}$ can be incorporated into $H_0$ by putting 
\begin{equation}\label{Kdef}
g_{c+} \equiv g = 
\left \{1+ 4\widetilde{V}_0(k\simeq 0)/\pi \hbar v_F \right\}^{-1/2}
 \leq 1 \;,
\end{equation}
while for all other channels, the coupling
constant $f$  gives rise to the tiny renormalization
$g_{a\neq c+}= 1+  f /\pi \hbar v_F \simeq 1$. 
The plasmon velocities of the four modes are 
$v_a=v_F/g_a$, and hence the charged $(c+)$ mode propagates with significantly
higher velocity than the three neutral modes. 

For the long-ranged interaction (\ref{unsc}), 
the logarithmic singularity in $\widetilde{V}_0(k)$
requires the infrared cutoff $k=2\pi/L$ due to the finite
length $L$ of the SWNT, resulting in:
\begin{equation} \label{longr}
g = \left\{ 1+\frac{8e^2}{\pi\kappa\hbar v_F} \ln(L/2\pi R) 
\right\}^{-1/2} \;.
\end{equation}
Since $\hbar c/e^2\simeq 137$, 
we  get with $v_F=8\times 10^5$ m/s the estimate
$e^2/\hbar v = (e^2/\hbar c) (c/v) \approx 2.7$, and
therefore $g$ is typically in the range $0.2$ to $0.3$. 
This estimate does only logarithmically depend on $L$ and $R$,
and should then apply to basically all SWNTs studied at the moment
(where $L/R\approx 10^3$).
The LL parameter $g$ predicted by Eq.~(\ref{longr}) 
can alternatively be written in the form
\begin{equation}
g = \left(1 + \frac{2E_c}{\Delta} \right)^{-\frac{1}{2}} \;,
\end{equation}	    
where $E_c$ is the charging energy and
$\Delta$ the single-particle level spacing. 
For our experimental setup described in Sec.~\ref{sec3}, 
the theoretically expected LL parameter is 
then estimated as $g_{\rm th} \approx 0.28$. 
The very small value of $g$  obtained here
implies that an individual metallic SWNT on an insulating substrate
is a strongly correlated system
displaying very pronounced non-Fermi liquid effects.

It is clear from Eqs.~(\ref{bh0}) and (\ref{bfs0}) that
for $f=b=0$, a SWNT constitutes a realization of
the LL. We therefore have to address the
effect of the nonlinear terms associated with the
coupling constants $f$ and $b$.  This can be done
by means of the renormalization group approach.
Together with a  solution via Majorana
refermionization,  this route allows for the complete
characterization of the non-Fermi-liquid ground state of a clean
nanotube \cite{egger2}.  $ $From this analysis, we find that for 
temperatures above the exponentially small energy gap 
\begin{equation}\label{massb}
k_B T_b = D \exp[-\pi \hbar v_F / \sqrt{2} b] 
\end{equation}
induced by electron-electron backscattering processes,
the SWNT is adequately described by the LL model, and $H_I^{(1,2)}$ 
can effectively be neglected.  A rough order-of-magnitude
estimate is $T_b \approx 0.1$ mK.  In the remainder, we focus on
temperatures well above $T_b$.

\subsection{Bulk and end tunneling: Scaling functions and exponents}
\label{sec22}

Under typical experimental conditions,  the contact between a SWNT and 
the attached (Fermi-liquid) leads is not perfect and
the conductance is limited by electron tunneling
into the SWNT, which in turn is governed by the
TDOS.  The TDOS exhibits power-law behavior and is strongly suppressed at
low energy scales.  The power-law exponent $\alpha>0$ depends
on the geometry of the particular experiment: If one tunnels
into the end of a SWNT, the exponent $\alpha_{\rm end}$ is
generally larger than the bulk exponent $\alpha_{\rm bulk}$,
since electrons can move in only one direction to
accomodate the incoming additional electron. 
The end-tunneling exponent can be easily obtained from
the open boundary bosonization technique \cite{bosonization}.
It follows that close to the boundary (taken at $x=0$),
i.e.~for $\max (x,x')\ll v_F t$, the single-electron Greens
function is of the form
\begin{equation}\label{OBB:GF}
\langle \Psi^{} (x,t)\Psi^\dagger(x',0)\rangle\propto t^{-(1/g+3)/4}\;.
\end{equation}
The boundary scaling dimension of the electron field
operator is therefore  
$\bar{\Delta}=\frac{1}{8g}+\frac{3}{8}$,
as opposed to its bulk scaling dimension
$\Delta=\frac{1}{16}\left(\frac{1}{g}+g\right)+\frac{3}{8}$.
Making use of the text-book definition of the TDOS
as the imaginary part of the electron Greens function,
we find from Eq.~(\ref{OBB:GF}) that the TDOS indeed vanishes as a power
law with energy,
\begin{equation} \label{tdos}
\rho(E)\propto (E/D)^\alpha \;,
\end{equation}
where the exponent $\alpha$ is given by the end-tunneling exponent
\begin{equation}\label{end}
\alpha_{\rm end}=2\bar{\Delta}-1= \left(\frac{1}{g} -1\right)/4 \;.
\end{equation}  
Similarly one may derive the bulk-tunneling exponent:
\begin{equation}\label{bulk}
\alpha_{\rm bulk}=2\Delta-1 = \left( \frac{1}{g}+g-2 \right)/8 \;.
\end{equation}
Since $\alpha >0$ for $g<1$,
the TDOS vanishes as the energy scale $E$
approaches zero in both cases. 
For a Fermi liquid, however, both exponents are zero.

If transport is limited by tunneling through a weak contact from a 
metal electrode to the SWNT, 
the full nonlinear and temperature-dependent differential conductance
$G(V,T)=dI/dV$ can be evaluated in closed form.  If $V$ denotes the voltage
drop across the weak link, one obtains 
\begin{equation}\label{scal}
G(V,T) = AT^\alpha \cosh\left(\frac{eV}{2k_BT}\right) \left|
\Gamma\left(\frac{1+\alpha}{2} + \frac{ieV}{2\pi k_B T}\right)\right|^2 \;,
\end{equation}
where $\Gamma$ denotes the gamma function
and $A$ is a nonuniversal prefactor depending on details of the junction.
The exponent $\alpha$ is either the end- or the bulk-tunneling exponent
depending on the experimental geometry.
If the leads are at finite temperature, the conductance is given by
a convolution of Eq.~(\ref{scal}) and the derivative of the Fermi function:
\[
-df/dE=\frac{1}{4k_B T\cosh^2(eV/2k_B T)} \;.
\]
Remarkably, the quantity $T^{-\alpha} G(V,T)$
should then be a {\sl universal}\, scaling function of the
 variable $eV/k_B T$ alone.
This scaling is seen experimentally as discussed  in Sec.~\ref{sec3}.

\subsection{Crossed nanotubes} 

\begin{figure}
\includegraphics[width=\textwidth]{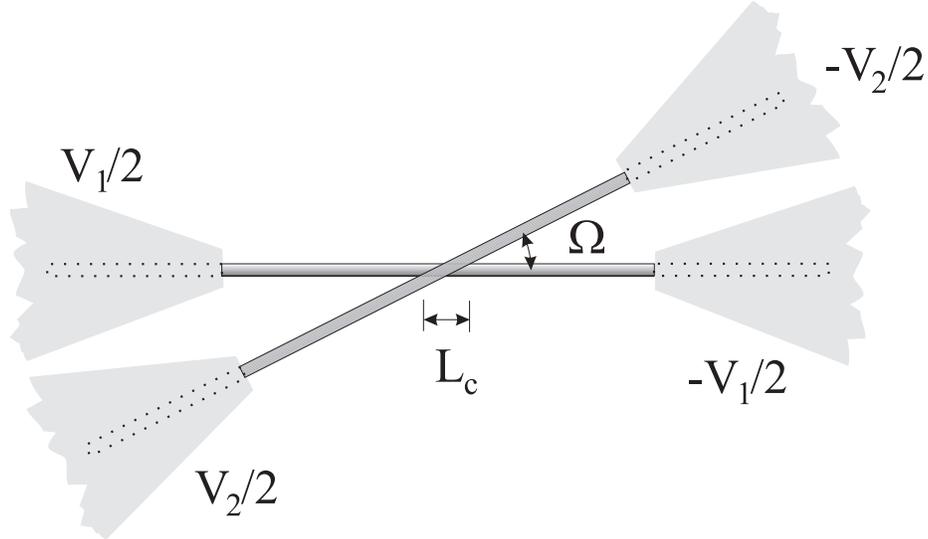}
\caption[]{\label{fig2}  Crossed nanotube setup.  By variation of 
the angle $\Omega$, the contact length $L_c$ can be changed. 
 We consider a pointlike contact, $L_c\leq a$.}
\end{figure}

More spectacular correlation effects can be observed in more
complicated geometries.  The simplest example is provided by
{\sl crossed nanotubes} \cite{komnik}
which have recently been studied experimentally \cite{kim,fuhrer}.
The geometry is shown in Figure \ref{fig2}, where we assume a 
pointlike contact of two clean metallic SWNTs characterized by the 
same $g$ parameter. 
External reservoirs can be incorporated by imposing Sommerfeld-like
radiative boundary conditions \cite{egger96}  
close to the contacts (for simplicity, 
we sketch the theory for the
spinless single-channel case).
This approach offers a general and 
powerful route to studying multi-terminal Landauer-B\"uttiker
geometries for correlated 1D systems.
Applying the two-terminal voltage $V_i$ along
conductor $i=1,2$, the boundary conditions read
\begin{equation}\label{bc}
\left(\frac{1}{g^2} \partial_x \pm \frac{1}{v_F} \partial_t\right) 
\langle \theta_i(x=\mp L/2, t) \rangle= \frac{eV_i}{\sqrt{\pi}\hbar v_F} \;.
\end{equation}
These boundary conditions fix the average densities of injected particles.
Outgoing particles are assumed to enter the reservoirs without reflection.

Let us now consider a point-like coupling at,
say, $x=0$.  Such a contact causes (at least) two different coupling
mechanisms.  First, there arises an
{\sl electrostatic interaction} 
$H_c^{(1)} \propto \rho_1(0) \rho_2(0)$. 
Bosonization shows that the only important part is 
\begin{equation}\label{v1}
H_c^{(1)} = \lambda \cos[\sqrt{4\pi} \,\theta_1(0)]
\cos[\sqrt{4\pi} \,\theta_2(0)] \;,
\end{equation}
which becomes relevant for sufficiently strong interactions, $g<1/2$.
The second potentially important process is  
{\sl single-electron tunneling} from one 
conductor into the other.  Notably, tunneling is
 always irrelevant for $g<1$,
and (unless the contact is very good) can therefore be treated in 
perturbation theory.
In other words, tunneling is expected to have only a
 very minor effect here, and
we shall hence focus on the effect of $H_c^{(1)}$ specified in 
Eq.~(\ref{v1}).
Again, for $g>1/2$, this term can also be treated 
perturbatively, but for the interesting
strong-interaction case $g<1/2$, qualitatively new
 features in the transport emerge.

\begin{figure}
\includegraphics[width=\textwidth]{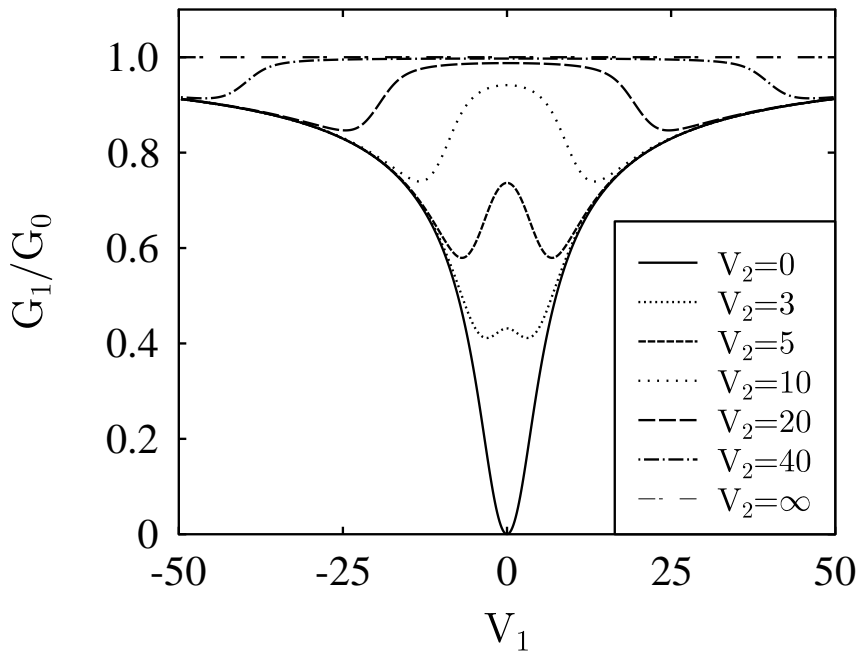}
\caption[]{\label{fig3} Conductance $G_1/G_0\equiv I_1/(e^2 V_1/h)$  
for $g=1/4, T=0$, and several values of the cross voltage $V_2$.
The overall energy scale is set by the coupling $\lambda$.}
\end{figure}

To investigate this situation further,
we switch to the linear combinations
$\theta_\pm (x) =  [\theta_1(x) \pm \theta_2(x)] /\sqrt{2}$,
whence the hamiltonian decouples into
the sum $H_+ + H_-$ with
\begin{equation} \label{decoup}
H_\pm = \frac{\hbar v_F}{2} \int dx \left \{
\Pi_\pm^2 + \frac{1}{g^2} (\partial_x \theta_\pm)^2 \right\}
\pm (\lambda/2)\, \cos \left[
\sqrt{8\pi} \,\theta_\pm(0)\right]\;.
\end{equation}
Effective boundary conditions (\ref{bc}) 
for the fields $\theta_\pm$ are found by simply 
replacing  $V_{1,2}\to (V_1\pm V_2)/\sqrt{2}$.
Therefore we are left with two completely decoupled
systems, each of which is formally identical to the
problem of an elastic potential scatterer embedded
into a spinless LL with effectively doubled interaction
strength parameter $g'=2g$. 
The hamiltonian (\ref{decoup})
has been discussed previously by Kane and Fisher \cite{kf},
and the exact solution 
under the boundary condition (\ref{bc})
has recently been given by boundary conformal 
field theory methods \cite{saleur}.
This solution applies for arbitrary 
$g,T,V,$ and $\lambda$.

The conductance  $G_1=I_1/(e^2 V_1/h)$ for $g=1/4$ at 
zero temperature is plotted 
as a function of $V_1$ and $V_2$  in Fig.~\ref{fig3}. 
Contrary to what is found in the uncorrelated 
case, $G_1$ is extremely sensitive to both $V_1$ and
$V_2$ (in a Fermi liquid, $G_1$ is simply constant).
For $V_2=0$, transport becomes fully suppressed for 
$V_1\to 0$, with a $g$-dependent
perfect {\sl zero-bias anomaly} (ZBA).
Remarkably, there is a suppression of the current if $|V_1|=|V_2|$,
which is observed as a ``dip''  in $G_1(V_1)$ for fixed $V_2$.
This effect can be rationalized in terms of 
a partial dynamical pinning of charge
density waves in tube 1 due to commensurate
 charge density waves in tube 2.
The consequence is that the ZBA dip at $V_1=0$ is turned into a peak 
by increasing the cross voltage $V_2$.
The pronounced and nonlinear  sensitivity of $G_1(V_1,V_2)$ 
to $V_2$ is a distinct fingerprint for LL behavior.  
Qualitatively, all these features have been observed in
a very recent experiment by Kim {\sl et al.}~\cite{kim}
on crossed MWNTs. 

\subsection{Spin transport}

The ultimate hallmark of a LL is electron fractionalization
and spin-charge separation.  So far no unambiguous experimental 
verification of spin-charge separation in a LL has been 
published, and carbon nanotubes might offer the possibility
 to do so.  The standard approach via photoemission
is clearly not suitable here since one should work on a single SWNT.
Alternatively, a spin transport experiment will be described below that 
should reveal spin-charge separation in a clear manner \cite{balents}.
In such an experiment, one needs to measure the $I-V$ characteristics
of a SWNT in weak contact to two {\sl ferromagnetic} reservoirs, where
the angle $\phi$ between the ferromagnet magnetization directions 
$\hat{m}_{1}$
and $\hat{m}_2$, i.e.~$\cos\phi=\hat{m}_1\cdot \hat{m}_2$, can take
an arbitrary value $0\leq \phi\leq \pi$.  A corresponding experiment
for $\phi=0,\pi$ has recently been performed for a MWNT \cite{FM-tube}.

Spin transport has been studied in detail for Fermi liquids.
For the proposed geometry of a metal connected to ferromagnetic leads
via tunnel junctions, Brataas {\sl et al.} \cite{brataas} have computed
the $\phi$-dependence of the current.  
Assuming identical junction and ferromagnet parameters, they obtain
\begin{equation} \label{currs}
\frac{I(\phi)}{I(0)} = 1-P^2 \frac{\tan^2(\phi/2)}{\tan^2(\phi/2) + Y} \;,
\end{equation}
where the polarization $0\leq P\leq 1$ parametrizes the difference in the 
spin-dependent DOS of a ferromagnetic reservoir,
and  $Y\geq 1$ is related to the spin-mixing conductance \cite{brataas}.
The result (\ref{currs}) shows that for any $\phi>0$ the current
will be suppressed due to the spin accumulation effect \cite{prinz}.
The maximum suppression, namely by a factor $1-P^2$, occurs for antiparallel
magnetizations, $\phi=\pi$.

If one has spin-charge separation, detailed analysis \cite{balents}
shows that the current is still properly described by Eq.~(\ref{currs}),
though with two important differences.  First, the current $I(0)$ for
parallel magnetizations will carry the usual power-law suppression
factor $(V/D)^{\alpha/2}$, where $\alpha>0$ is the bulk/end tunneling
exponent.  More importantly, 
the quantity $Y$ will now be $V$- and $T$-dependent,
with a divergence as $V,T \to 0$ according to
$Y\propto [{\rm max}(eV,k_B T)/D]^{-\alpha}$.  
Therefore the spin accumulation
effect, i.e.~the suppression of the current by changing $\phi$
away from zero,
will be totally destroyed by spin-charge separation, except for $\phi=\pi$.
This qualitative difference to a Fermi liquid should be easily detectable
and can serve as a signature of spin-charge separation.

\section{Experimental evidence for Luttinger liquid}
\label{sec3}

In this section, we show first 
with electrostatic force microscopy (EFM) that metallic nanotubes are ballistic
conductors, an important ingredient for the possible observation of LL behavior.
When nanotubes are attached to metallic electrodes, EFM shows that
a barrier is formed at the nanotube/metal interface. This fact is then exploited
to observe LL behavior in nanotube devices via the TDOS.
We show experimentally that the TDOS indeed exhibits 
power-law behavior in metallic SWNTs. 
This is observed by mesuring the tunneling
conductance of nanotube/metal interfaces as a function of temperature and
voltage. 

\subsection{Electrostatic Force Microscopy of electronic transport 
in carbon nanotubes}\label{EFM}

\begin{figure}
\includegraphics[width=\textwidth]{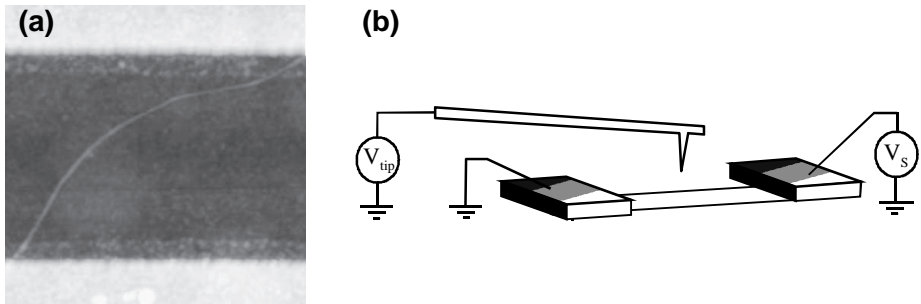}
\caption[]{\label{fig4} (a) Topographic AFM image of a 
\mbox{2.5~nm} diameter bundle of SWNTs which is seen spanning 
between two gold electrodes. 
The separation between the electrodes is 1~$\mu$m. 
(b) Experimental setup for EFM. A conducting AFM cantilever 
is scanned above the device, which consists of a nanotube 
contacted by two gold electrodes. Adapted from Ref.~\cite{bachtold}.}
\end{figure}

Samples are fabricated on a backgated substrate consisting 
of degenerately doped silicon capped with 1~$\mu$m SiO$_2$. 
SWNTs synthesized via laser ablation are ultrasonically 
suspended in dichloroethane, and the resulting suspension 
is placed on the substrate for approximately 15 seconds,
then washed off with isopropanol. An array of structure, 
each consisting of two Cr/Au electrodes, is fabricated using 
electron beam lithography. Samples that have a measurable 
resistance between the electrodes are selected with a prober. 
An AFM is then used to choose samples 
that have only one nanotube rope between the electrodes. 
Objects whose height profile is consistent with single SWNTs 
(1-2~nm) are preferentially selected. An example of a SWNT 
rope contacted by two electrodes is shown in Fig.~\ref{fig4}(a). 
Since the success of this contacting scheme works by chance,
 it is obvious that the yield is low. However, since a large array of 
structures can readily be fabricated, this scheme has
 turned out to be very convenient.

We continue by reviewing the EFM technique \cite{efm2} 
which is used to directly probe the nature of conduction in SWNTs.  
An AFM tip with a voltage $V_{\rm tip}$ is scanned over a 
nanotube sample, see Fig.~\ref{fig4}(b). The electrostatic force between the tip
 and the sample is given by
\begin{equation}
F = \frac{1}{2}\frac{dC}{dz}(V_{\rm tip} +  \phi - V_s)^2 \;,
\end{equation}
where $V_s$ is the voltage within the sample, $\phi$ is the work 
function difference between the tip and sample, and $C$ is the
tip-sample capacitance. The tip is held at constant height above 
the surface by first making a line-scan of the topography of the 
surface using intermittent-contact AFM, and then making a second 
pass with the tip held at a fixed distance above the measured 
topographic features. In order to detect the electrostatic force,
 the cantilever is made to oscillate by an AC potential that is 
applied to the sample at the resonant frequency of the cantilever. 
 This produces an AC force on the cantilever proportional to the 
local AC potential $V_s(w)$ beneath the tip:  
\begin{equation}
F_{ac}(w) = \frac{dC}{dz}(V_{\rm tip} + \phi) \, V_s(w) \;.
\end{equation}
The resulting oscillation amplitude is recorded using an
external lock-in amplifier; the signal is proportional
 to $V_s(w)$. Calibration of this signal is made by 
applying a uniform $V_s(w)$ to the whole sample and measuring 
the response of the cantilever.

EFM yields a signal that is proportional to the local 
voltage within the nanotube circuit.  However, the signal 
is also proportional to the derivative of the local capacitance. 
 This will vary as the geometry changes, yielding e.g.~different 
signals over a nanotube than over a contact at the same potential.  
However, $dC/dz$ does not vary appreciably as a function of distance 
along the nanotube.  The measured signal should thus accurately 
reflect the local voltage within the nanotube.

\subsection{Ballistic transport in metallic SWNTs}

Next we discuss measurements of the device shown in Fig.~\ref{fig4}.
The resistance of this 2.5~nm diameter bundle 
is 40~k$\Omega$ and has no significant gate voltage dependence. 
We have also measured the current at large biases -- 
the current saturates at 50~$\mu$A.  
This  is in agreement with recent work by Yao, Kane and Dekker \cite{Yao}
 where the current was observed to be limited to 25~$\mu$A 
per metallic nanotube due to optical or zone-boundary 
phonon scattering.  We therefore conclude that the current 
is carried by 2 metallic SWNTs in the bundle.

\begin{figure}
\includegraphics[width=\textwidth]{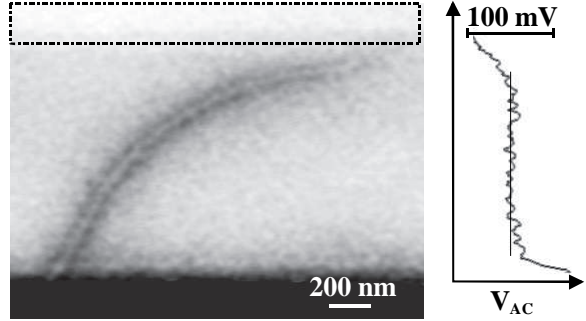}
\caption[]{\label{fig5} EFM image of the same bundle of SWNTs shown 
in Fig.~\ref{fig4}. An AC potential of 100~mV is applied to the 
lower electrode. The upper electrode indicated by the box is grounded. 
The AC-EFM signal is flat along the length of the SWNT bundle, 
indicating that the potential drops occur at the contacts, 
and not along the bundle length.  A trace of the potential as a 
function of vertical position in the image is also shown. 
Adapted from Ref.~\cite{bachtold}.}
\end{figure}

Figure \ref{fig5} shows the EFM image of this SWNT bundle, 
as well as a line trace along the backbone of the bundle. 
The potential is flat over its length, indicating that within our 
measurement accuracy there is no measurable intrinsic resistance.  
Taking into account the finite measurement resolution, 
we estimate that $R_i$ of the bundle is at most 3~k$\Omega$.
The contact resistances are measured to be approximately 28~k$\Omega$ 
and 12~k$\Omega$ for the upper and lower contacts, respectively. 
The conductance of the tube has been controlled to no change
 when the tip scans over it. The original data have a background 
signal due to stray capacitive coupling of the tip to the
 large metal electrodes. The image in Fig.~\ref{fig5} is shown 
with the background signal subtracted according to the
 procedure described in Ref.~\cite{bachtold}. 

Using the four-terminal Landauer formula,
 $R = (h/4e^2)(1-T_i)/T_i$  per nanotube, where $T_i$ is 
the transmission coefficient for electrons along the length of the 
nanotube, we find that $T_i$ is larger than 0.5.  
This indicates that the majority of electrons are transported 
through the bundle with no scattering.
Therefore transport is ballistic at room temperature over
 a length of $>1\mu$m.  
This confirms the theoretical  predictions of very weak 
scattering in metallic SWNTs \cite{white,anantram}. 
This is also in agreement with previous low-temperature 
transport measurements which indicate that long metallic SWNTs 
may behave as single quantum dots  \cite{tans,bockrath2}, 
and room-temperature measurements of metallic 
SWNTs which sometimes exhibit low two-terminal resistance \cite{soh}.  
The dominant portion of the overall resistance of 40~k$\Omega$ 
thus comes from the contacts, indicating that
the transmission coefficients 
for entering and leaving the bundle are significantly 
less than one and the contacts are not ideal.  
As discussed in the next section,
this fact can be exploited to observe LL behavior 
in nanotube devices via the tunneling density of states (TDOS).

\subsection{Tunneling conductance}

\begin{figure}
\includegraphics[width=\textwidth]{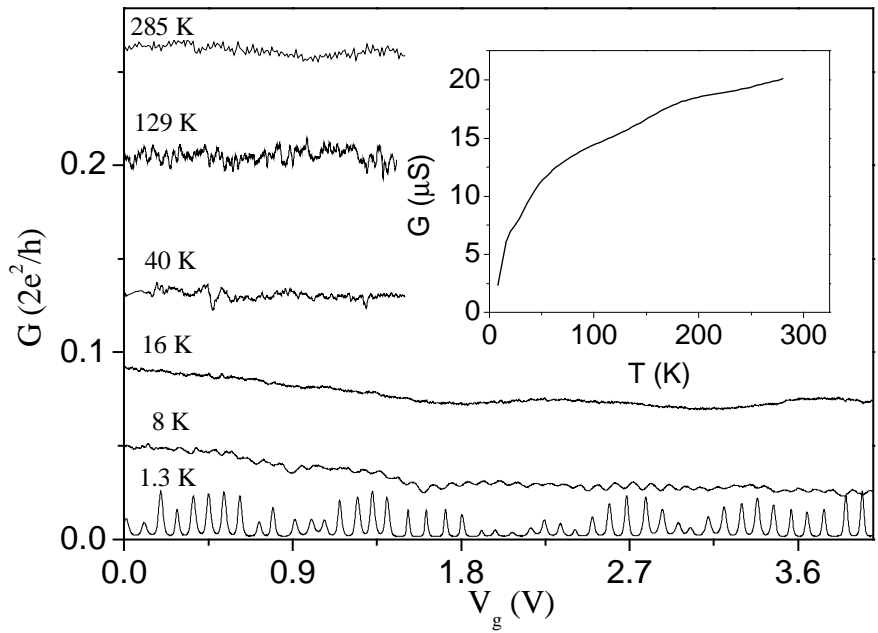}
\caption[]{\label{fig6} The two-terminal linear-reponse conductance
 as a function of gate voltage at a variety of temperatures. 
The inset shows the average conductance as a function of temperature.
 Adapted from Ref.~\cite{bockrath}.}
\end{figure}

The fact that a metallic nanotube acts like a nearly perfect
1D conductor with very long mean free path makes it an ideal 
system to test the LL theory described in Sec.~\ref{sec2}.
Figure \ref{fig6} shows the linear-response 
two-terminal conductance $G$ versus gate voltage $V_g$ 
for a metallic rope at different temperatures. 
At low temperatures, the conductance exhibits a series of
 Coulomb oscillations with a charging energy $E_c= 1.9$~meV.  
For $k_B T > E_c$, i.e. $T > 20 K$, the Coulomb oscillations are
nearly completely washed out, and the conductance is 
independent of gate voltage.  A plot of the conductance vs.~temperature 
in this regime is shown in the inset. The conductance drops 
steeply as the temperature is lowered, extrapolating to 
$G=0$ at  $T=0$.

\begin{figure}
\includegraphics[width=\textwidth]{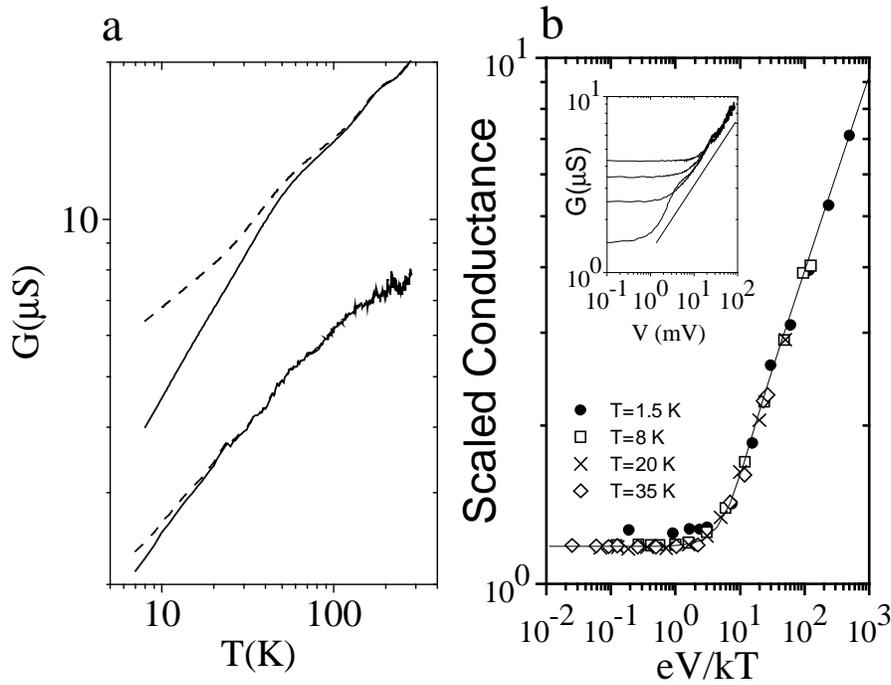}
\caption[]{\label{fig7} (a) Conductance plotted against 
temperature on a double-logarithmic
 scale for two samples. (b) The scaled 
differential conductance measured at different temperatures. 
Adapted from Ref.~\cite{bockrath}.}
\end{figure}

Results for two samples are shown in Fig.~\ref{fig7}(a), 
where the conductance as a function of temperature is 
plotted on a double-logarithmic scale (solid curves). 
Charging effects contribute to the measured characteristics, 
especially at lower temperatures, $k_B T < 2E_c$. 
We therefore correct the $G(T)$ data for charging  effects by
 dividing the measured conductance by the theoretically expected
 temperature dependence of $G$ in the Coulomb blockade model \cite{grabert}. 
The dashed lines in Fig.~\ref{fig7}(a) show the measured $G$ corrected in this 
manner as a function of temperature. Looking at the corrected data, 
we see that they have a finite slope, indicating an 
approximate power-law dependence upon temperature with 
exponents $\alpha=0.33$ and 0.38. 

Figure \ref{fig7}(b) shows the measured differential 
conductance as a function of the applied bias $V$.  
The upper left inset to Fig.~\ref{fig7}(b) shows $G=dI/dV$ 
versus $V$ at different temperatures, plotted on a double-logarithmic
scale.  At low bias, $dI/dV$ is proportional to a 
(temperature-dependent) constant. At high bias, $dI/dV$ 
increases with increasing $V$. The curves at different temperatures 
fall onto a single curve in the high-bias regime. 
Since this curve is roughly linear on the double-logarithmic
plot,  the differential conductance
is well described by a power law, $dI/dV \propto V^{\alpha}$, 
where $\alpha = 0.36$.
 At the lowest temperature $T=1.6$~K, 
this power-law behavior extends over two decades in the applied voltage $V$, 
namely from 1~mV up to 100~mV.

Let us next discuss possible origins of this behavior.
 The data demonstrates that tunneling into the rope has a 
significant dependence on energy that cannot be described by the 
Coulomb blockade model. 
 One simple explanation for such behavior is that the transmission 
coefficients of tunnel barriers are strongly energy-dependent, with 
substantially increased transparency on the energy scale of the 
measurement.  This would lead e.g.~to activated transport over the barrier,
$G \propto \exp(-\Delta/k_B T)$.  
However, the fact that the temperature dependence
 extra\-polates to \mbox{$G$ = 0} at \mbox{$T$ = 0}, see inset of
Fig.~\ref{fig6}, is inconsistent with this functional form. 
The origin of this behavior appears to originate rather 
from the TDOS of a LL which vanishes as a 
power law with energy, see Eq.~(\ref{tdos}).
The assumption that the conductance is limited by tunneling between
the metal electrodes and the LL
directly leads to the power-law temperature dependence 
$G(T) \propto T^{\alpha}$ at small bias, $eV \ll k_B T$.
Similarly, for $eV\gg k_B T$, LL theory predicts that
$G(V) \propto V^{\alpha}$.
The exponent $\alpha$ follows from the corresponding 
(bulk-tunneling, see below) exponent in the TDOS. We therefore obtain
$\alpha_{\rm bulk} \approx 0.3$.

The devices used here were made in the following way. 
Electron beam lithography is first used to define leads, 
and ropes are deposited on top of the leads. 
Samples were selected that showed Coulomb blockade behavior 
at low temperatures with a single well-defined period,
 indicating the presence of a single dot.
 The charging energy of these samples indicates a dot with a size 
substantially larger than the spacing between the leads \cite{tans}. 
Transport thus occurs by electrons tunneling into the middle (``bulk'')
of the nanotubes. These devices are referred to as ``bulk-contacted.''
We can also use a second method, which was described in Sec.~\ref{EFM}, 
where the contacts are applied over the top of the nanotube rope.
$ $From measurements of these devices in the Coulomb blockade regime
 \cite{bockrath2}, we conclude that the electrons are confined to the 
length of the rope between the leads. This implies that the 
leads cut the nanotubes into segments, and transport 
involves tunneling into the ends of the nanotubes. This type of device 
is referred to as ``end-contacted.''
For end-contacted devices, similar temperature and voltage 
dependences of the conductance $G(V,T)$ are observed.
The obtained exponent $\alpha_{\rm end}\approx 0.6$ is
significantly larger than $\alpha_{\rm bulk}\approx 0.3$,
the exponent obtained for bulk-contacted devices.

The exponent of these power laws obviously depends on whether the 
electron tunnels into the end or the bulk of the LL. 
These exponents are related to the LL parameter $g$ 
by Eq.~(\ref{end}) and (\ref{bulk}), respectively.
Using the expected LL parameter $g_{\rm th}=0.28$, 
see Sec.~\ref{sec22}, the expected exponents are 
$\alpha_{\rm end, th} = 0.65$ and $\alpha_{\rm bulk, th}= 0.24$. 
The approximate power-law behavior as a function of $T$ or $V$
observed in Fig.~\ref{fig7} follows the theory 
for tunneling into a LL. The predicted values of the exponents 
are in good agreement with the experimental values. 
Remarkably, power-law behavior in $T$ is observed up to 300~K, 
indicating that nanotubes are LLs even at room temperature.

LL theory makes an additional prediction for this system. 
Since the temperature and the voltage play an analogous role in the 
theory, the differential conductance for a single tunnel junction 
should obey the universal scaling form (\ref{scal}), together with a 
convolution of the derivative of the Fermi
 distribution, see Sec.~\ref{sec22}. 
Hence it should be possible to collapse the data onto a single universal 
scaling curve. To do this, the measured nonlinear conductance 
$G(V,T)=dI/dV$ at each temperature was divided 
by $T^{\alpha}$ and plotted against $eV/k_B T$, 
as shown in the main body of Fig.~\ref{fig7}(b). 
The data collapses well onto a universal curve. 
The solid line in Fig.~\ref{fig7}(b) is the theoretical plot,
see Ref.~\cite{bockrath} for details. 
The theory fits the scaled data quite well.

Recently, Yao {\sl et al.}~\cite{yao} have reported on
electrical transport measurements 
on SWNTs with intramolecular junctions. Two nanotubes are
 connected together by a kink, which acts as a tunnel barrier.
 In the case of a metal-metal junction, the conductance 
displays a power-law dependence on temperatures and voltage, 
consistent with tunneling between the ends of two LLs.
 The tunneling conductance $G$ across the junction is 
proportional to the product of the end-tunneling DOS on both sides.
Therefore $G$ still varies as a power law of energy,
 but with an exponent twice as large, namely
$\alpha_{\rm end-end} = 2\alpha_{\rm end}$.

\section{Discussion and open problems} 
\label{sec4}

In this review, we have discussed our recent observation of 
Luttinger liquid behavior in transport experiments on 
individual metallic  carbon nanotubes, along
with the detailed theoretical description of this non-Fermi liquid
state.  The situation
in SWNTs seems rather clear by now, since the
ballistic nature of transport can be unambiguously established.
Nevertheless, several interesting open questions remain.
One proposed experiment could probe spin-charge separation
by measuring the $I-V$ characteristics
of a SWNT in contact to two ferromagnetic reservoirs with continuously
varying angle $\phi$ between the magnetization directions.
Another interesting issue concerns the experimental observation 
of Friedel oscillations in nanotubes, i.e.~density oscillations 
in the conduction electron density around impurities or the end
of the tube.  These density oscillations should decay with a
slow interaction-dependent power law (slower than $1/x$), and,
interestingly, the oscillation period can depend on the 
interaction strength \cite{egger2}. 
Furthermore, it is of importance to achieve a better
understanding of conduction electron spin resonance (CESR) in SWNTs.
Previous experimental attempts have not seen any ESR peak,
and one of the proposed reasons for its absence
involves electron-electron interactions \cite{forr}. 
However, to the best of our knowledge, there are no theoretical
investigations concerning CESR for Luttinger liquids including the
gapless charge degrees of freedom. 
Finally, the phonon backscattering correction to the conductance arising for long SWNTs 
at high temperatures should involve an anomalous $T^{(1+g)/2}$ 
scaling \cite{komnik2} that remains to be seen experimentally.

Another line of research currently deals with multi-wall nanotubes
which are known to exhibit diffusive transport.  Nevertheless, the
TDOS apparently shows very similar behaviors  as in a SWNT, and superficially
it appears that Luttinger liquid concepts also apply to MWNTs.
The reason for this is presently unclear, and more
theoretical and experimental studies will be needed to clarify the
situation.

We acknowledges support by the DFG under the Gerhard-Hess program,
by the DOE (Basic Energy Sciences,
 Materials Sciences Division, the sp2 Materials Initiative), 
and by DARPA (Moletronics Initiative).

\end{document}